\documentclass[iop]{emulateapj}

\shorttitle{The $\lambda$10830 \ion{He}{1} Line Among Subdwarfs}
\shortauthors{Smith et al.}

\begin{document}

\title{The $\lambda$108300 \ion{He}{1} Absorption Line Among Metal-Poor 
Subdwarfs\footnote{The data presented herein were obtained at the W.M. Keck 
Observatory, which is operated as a scientific partnership among the California
Institute of Technology, the University of California and the National 
Aeronautics and Space Administration. The Observatory was made possible by the 
generous financial support of the W.M. Keck Foundation.}} 

\author{Graeme H. Smith}
\affil{University of California Observatories\/Lick Observatory, Department 
 of Astronomy \& Astrophysics, UC Santa Cruz, 1156 High St., Santa Cruz, CA 
 95064, USA}
\email{graeme@ucolick.org}

\author{Andrea K. Dupree}
\affil{Harvard-Smithsonian Center for Astrophysics, Cambridge, MA 02138, USA}
\email{dupree@cfa.harvard.edu}

\author{Jay Strader}
\affil{Department of Physics and Astronomy, Michigan State University, 
 East Lansing, MI 48824, USA}
\email{strader@pa.msu.edu}

\slugcomment{Published in PASP}

\begin{abstract}
Spectra of the \ion{He}{1} $\lambda$10830 line have been obtained for 23
metal-poor stars, the majority of which are dwarfs ranging in metallicity from
$-2.1 \leq {\rm [Fe/H]} \leq -0.8$. The data were acquired with the NIRSPEC
spectrograph on the Keck 2 telescope. Most of these subdwarfs and dwarfs are 
found to exhibit a \ion{He}{1} absorption line indicative of the presence of 
chromospheres. The equivalent width of the $\lambda$10830 absorption profile 
is generally less than 70 m\AA, and covers a range similar to that found in 
solar metallicity stars of low activity. Among the subdwarfs the $\lambda$10830
equivalent width does not correlate with either [Fe/H] metallicity or $(B-V)$ 
color. Some evidence for asymmetric profiles is found among metal-poor dwarfs, 
but not the high-speed blue-shifted absorption displayed by some metal-poor 
red giants.
\end{abstract}

\keywords{Stars} 

\section{Introduction}

Whereas chromospheres are known to be present among both metal-poor 
Population II subdwarfs (Peterson \& Schrijver 1997; Smith \& Churchill 1998;
Takeda \& Takada-Hidai 2011) and giants (Dupree et al. 1990, 2007; Dupree \& 
Smith 1995; Smith \& Dupree  1998; Cacciari et al. 2004; Mauas et al. 2006; 
M\'{e}sz\'{a}ros et al. 2009; Smith et al. 1992; Vieytes et al. 2011) 
despite their great age, 
and whereas the \ion{He}{1} triplet line at 10830 \AA\ can serve as a tracer of
chromospheric conditions among Luminosity Class V and III F-G-K stars (Zirin 
1982; O'Brien \& Lambert 1986; Zarro \& Zirin 1986; Shcherbakov \& Shcherbakova
1991; Andretta \& Giampapa 1995; Sanz-Forcada \& Dupree 2008), and reveal mass 
motions of quite high velocity among red giants (O'Brien \& Lambert 1986), 
and whereas the profiles of the \ion{He}{1} 10830 \AA\ line have been found to 
exhibit evidence of fast outflows among metal-poor red
giants (Dupree et al. 1992, 2009; Smith et al. 2004), it seems warranted to 
document the behavior of the \ion{He}{1} 10830 \AA\ profile among metal-poor
dwarf and subdwarf stars. If the fast outflows found among some 
metal-poor red giant branch (RGB) and red horizontal branch (RHB) stars
(Dupree et al. 2009) are to be associated with mass loss that uniquely
occurs during the red giant phase of evolution then one might not expect 
evidence of them among subdwarf stars. On the other hand, if the \ion{He}{1}
fast outflows carry very minor amounts of mass and prove to be commonplace 
among the upper atmospheres of late-type stars of widely differing surface
gravity and luminosity, then it may be concluded that
they are of no significance in the mass loss history of Population II red 
giants. It is to address this issue that we report upon below the results of 
a spectroscopic survey of the \ion{He}{1} line for a sample of metal-poor dwarf
stars that was carried out with the NIRSPEC spectrometer on the Keck 2
telescope. A comparable program of similar scope has previously been
published by Takeda \& Takada-Hidai (2011) using the Subaru telescope. Our 
surveys prove to be largely complementary and there is only modest overlap 
in the stars observed in these two programs.

\section{Observations}

A selection of metal-poor halo dwarf stars was observed over the course of
$1\frac{1}{2}$ nights in November 2009 using the NIRSPEC spectrograph 
(McLean et al. 2000) on the 
Keck 2 telescope. The stars reported upon in this paper were chosen from the 
{\it Catalogue of [Fe/H] Determinations of F, G, K stars} compiled by 
Cayrel de Strobel et al. (2001). They were selected from among those stars for 
which at least one published abundance measurement yielded a result of 
${\rm [Fe/H]} < -0.8$. The stars were also listed as being of luminosity class 
V or IV in the Cayrel de Strobel et al. (2001) catalogue. 

The program stars are listed in Table 1. Photometry in columns 3 and 4 of the 
table is taken from {\it The General Catalogue of Photometric Data} (Mermilliod
et al. 1997) where available, for a few stars these values are taken from the 
{\it Hipparcos Catalog} (ESA 1997). Parallaxes (column 5) in most cases are 
also taken from the Hipparcos Catalog, but from Lepine (2007) in the case of 
HD 23439B. Absolute magnitudes in column 6 have been computed without any 
attempt to correct for possible instellar absorption along the line of sight.
Spectroscopic metallicities for each star are listed in Table 1 (column 7) 
together with a note of the references (column 8) from which the values of 
[Fe/H] were obtained. In several cases where there is some dispersion in the 
literature measurements two values of [Fe/H] are listed. Radial velocities 
$v_r$ as taken from the SIMBAD data base are listed in column 9 of Table 1. 
Of the stars in our sample HD 194598 and HD 201891 were also observed by 
Takeda \& Takada-Hidai (2011).

Observations were made on the night of 2009 November 22 UT and 
the first half of the night of November 23 UT. NIRSPEC 
(McLean et al. 1998) was used in high-resolution mode with the NIRSPEC-1 
above-slit filter and echelle and cross-disperser angles chosen so as to place 
the wavelength range from 0.95 to 1.12 $\mu$m on the ALADDIN-3 1024$\times$1024
detector (which has 27 $\mu$m pixels). The 1.083 $\mu$m 
\ion{He}{1} line is located within order 70. The thin blocking filter which is
available with NIRSPEC to reduce longer-wavelength thermal leakage was not
used in these observations due to concern that it would introduce fringing into
the spectra on wavelength scales comparable to the width of the stellar 
\ion{He}{1} line. A 0.43$\times$12 arsec slit was used in all observations.

The total exposure time allocated to each star is listed in column 2 of Table 1
as the product of three factors that define how each observation was carried 
out. The first digit is the number of nod points at which integrations were 
performed, it is either 2 (for an AB nod pattern) or 4 (for an ABBA pattern). 
The second digit is the number of coadded integrations made at each nod 
position and the third digit gives the integration time in seconds per coadd. 
The only star that was observed with an ABBA nod pattern is HD 194598, the 
other stars in Table 1 were observed in an AB nod pair. In the instances of 
HD 31128 and HD 76932 two AB nod pairs were performed, and the integration 
times per coadd for each pair are listed in brackets.

Observations of a variety of hot rapidly-rotating bright stars were 
interspersed throughout the night for monitoring telluric absorption features.
Spectra of flat field lamps and NeArKr comparison arcs were also obtained at 
the beginning and end of 22 November, and at the beginning of the 23 November
half-night. A spectrum of a K giant was used to verify the dispersion 
relation.

\section{Results}

The ($M_V$, $B-V$) color-magnitude diagram (CMD) of the stars from Table 1 is 
shown in Figure 1. The $B-V$ colors plotted are the observed colors, no 
correction has been made for interstellar reddening. 
As a precedent, Gratton et al. (2000) assumed that metal-poor dwarfs in 
their abundance survey, which includes a number of stars in Table 1, had 
negligible reddening. They considered this assumption to be reasonable
``in view of their small distances.'' A large fraction of the stars in Table 1
are included in a re-analysis of the Geneva-Copenhagen survey of solar 
neighborhood dwarf stars performed by Casagrande et al. (2011). For most
of these stars the reddening in their catalog is quoted as $E(B-V) = 0.00$.
Only in one case is a small non-zero reddening given by them, namely, 
HD 97916 ($E(B-V) = 0.025$).\footnote{Stars from Table 1 not included in 
Casagrande et al. (2011) are HD 3567, HD 7424, HD 16031, HD 23439B, HD 194598.}
Those program stars with $M_V > +3.0$ fall at locations in the CMD that are 
consistent with them being dwarfs. HD 3179 with $M_V \sim +2.5$ falls near
the base of the red giant branch and could be classified as a subgiant.

\begin{figure}
\figurenum{1}
\begin{center}
\includegraphics[angle=90,scale=0.4]{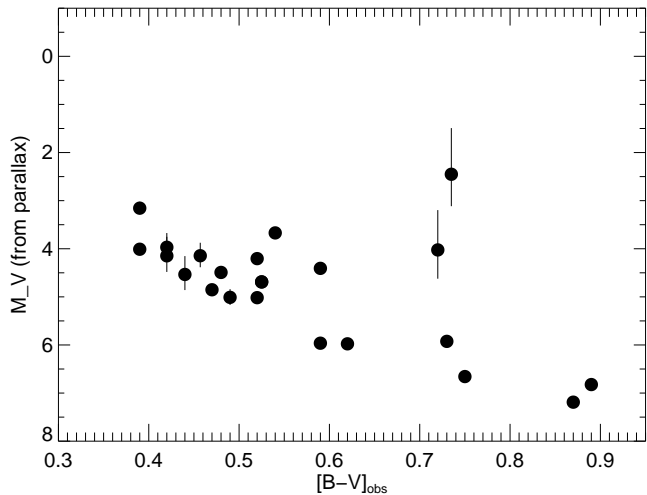}
\caption{The ($M_V$, $B-V$) color-magnitude diagram of stars with \ion{He}{1} 
 spectra listed in Table 1. The error bars on the absolute magnitude are 
 derived from the uncertainties in the {\it Hipparcos} parallaxes. In most
 cases the error bars are smaller than the diameter of the plotted points.}
\end{center}
\end{figure}

The spectra were reduced using the IDL package 
REDSPEC\footnote{\tt http://www2.keck.hawaii.edu/inst/nirspec/redspec.html}
McLean et al. (2003), as described in Dupree et al. (2009). 
Spectra were normalized by a continuum that was derived using a cubic spline 
fit to the line-free portions of the data.

The spectrum of each star in Table 1 derived from the NIRSPEC order containing 
the $\lambda$10830 \ion{He}{1} line is shown in one of the various panels of 
Figure 2(a-b). The wavelength scale of each panel corresponds to the rest frame
of the metal-poor star plotted in that panel. Several rapidly rotating hot 
stars were observed throughout the night to document the pattern of telluric 
water vapor absorption lines in the NIRSPEC order containing the stellar 
\ion{He}{1} line. One of these hot star spectra is plotted in each of the 
panels of Figure 2, shifted in wavelength so as to match the rest frame of 
each star. Comparison between the metal-poor and hot star spectra in each panel
allow the telluric lines to be identified in each metal-poor star spectrum. 

\begin{figure*}[!hp]
\figurenum{2a}
\begin{center}
\includegraphics[angle=90,scale=0.57]{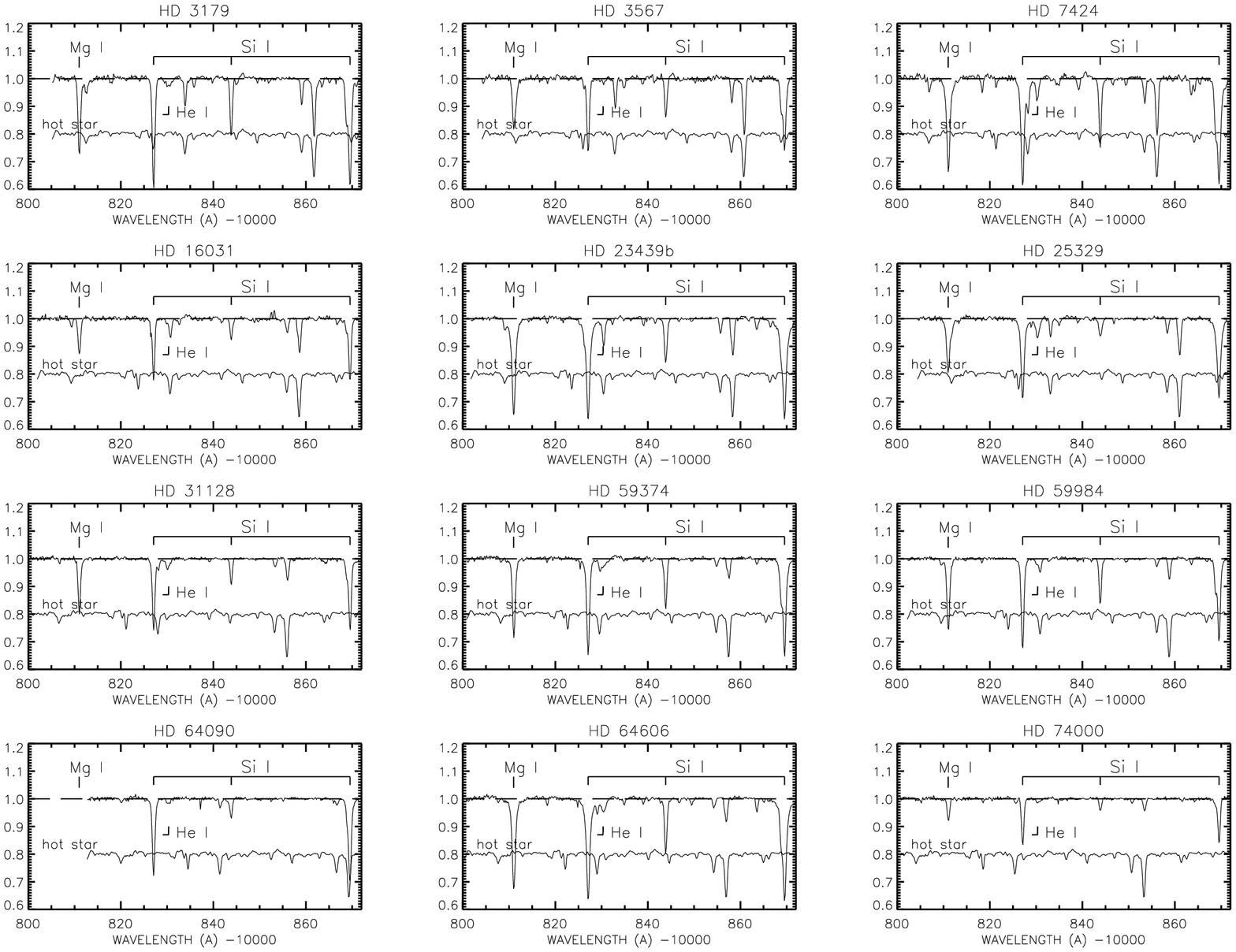}
\caption{Figure 2a. See caption below.}
\end{center}
\end{figure*}

\begin{figure*}[!hp]
\figurenum{2b}
\begin{center}
\includegraphics[angle=90,scale=0.57]{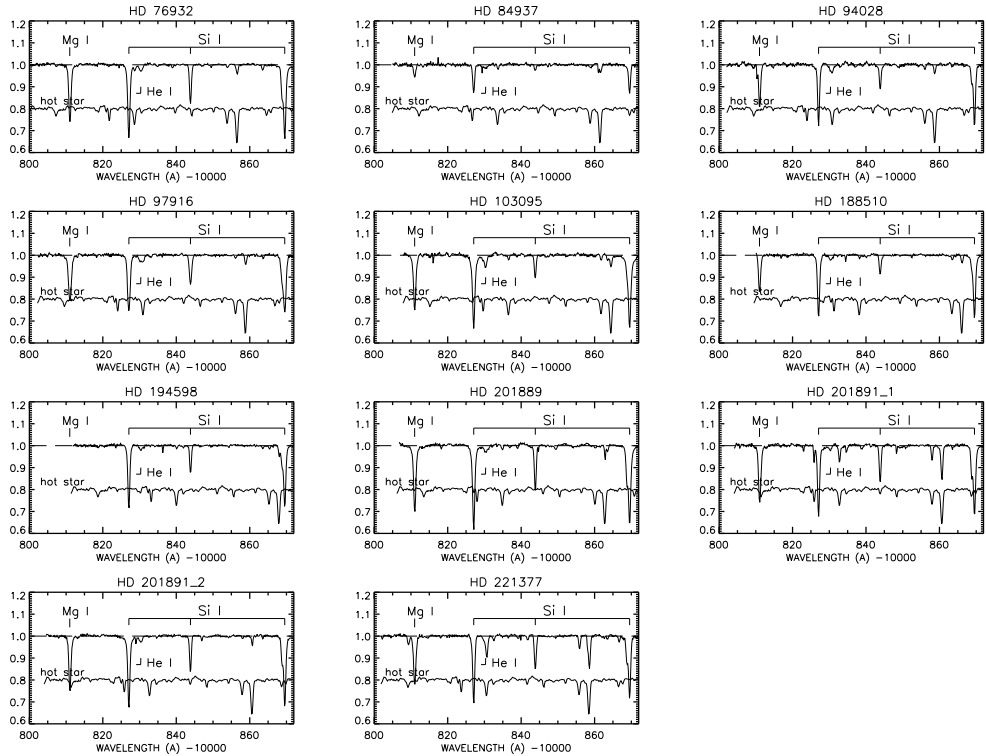}
\caption{Spectra of the wavelength range $\lambda$10800-10870 for metal-poor stars listed in Table 1, with the stellar absorption lines of \ion{He}{1}, \ion{Mg}{1} and \ion{Si}{1} marked. In each panel the spectrum of a  hot star is also shown that is shifted in wavelength so that the telluric lines match the positions of those in the subdwarf spectrum. The `hot star' shows a typical water vapor spectrum and is the same for all  panels, it is not scaled to the amount of water vapor in the spectrum of the  metal-poor star in each panel. All spectra have been normalized by the local continuum, and the hot star spectrum is shifted vertically by an arbitrary  amount. The wavelength scale is that of the rest frame of the metal-poor star.}
\end{center}
\end{figure*}

\begin{figure*}[!hp]
\figurenum{3a}
\begin{center}
\includegraphics[angle=90,scale=0.60]{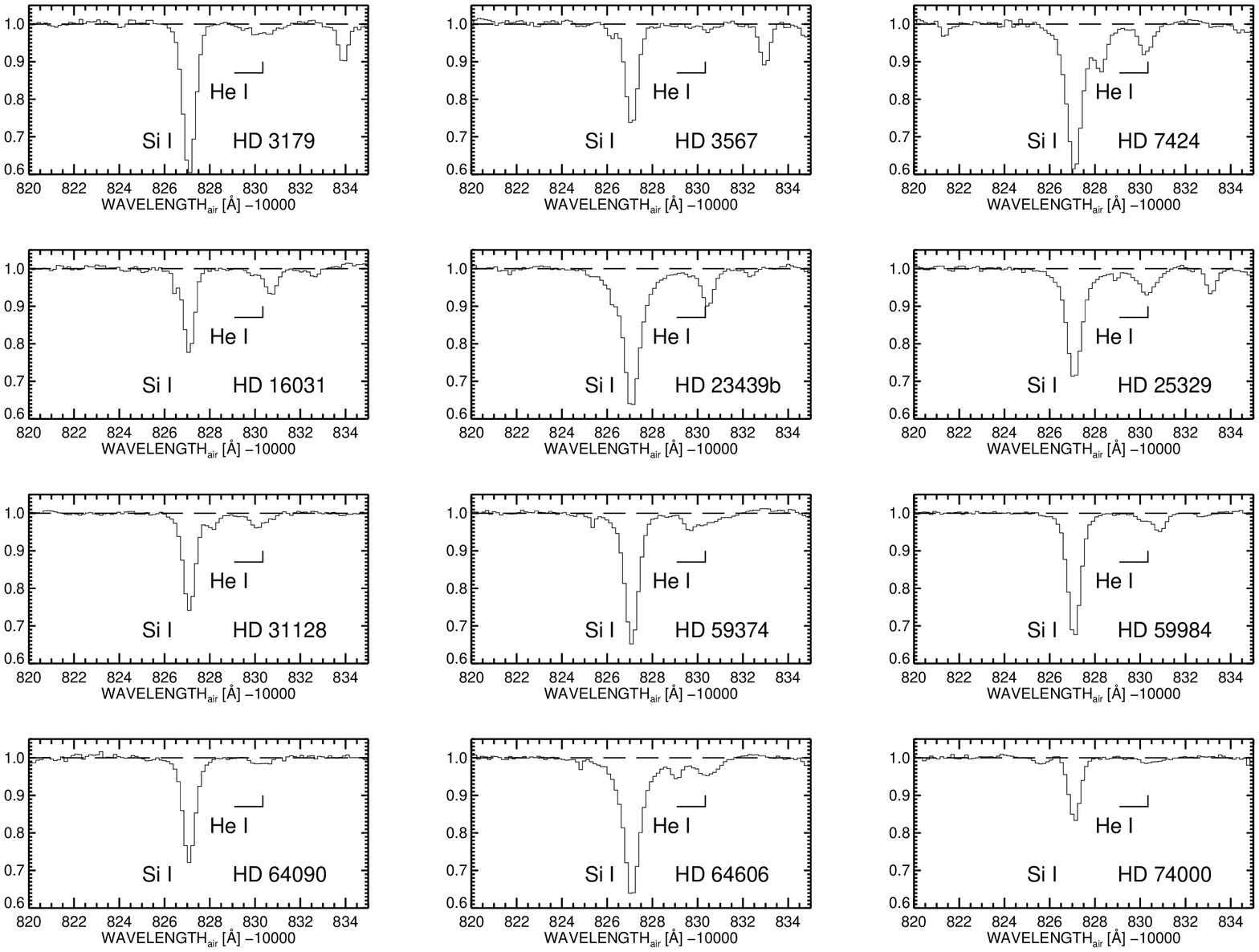}
\caption{Figure 3a. See caption below.}
\end{center}
\end{figure*}

\begin{figure*}[!hp]
\figurenum{3b}
\begin{center}
\includegraphics[angle=90,scale=0.60]{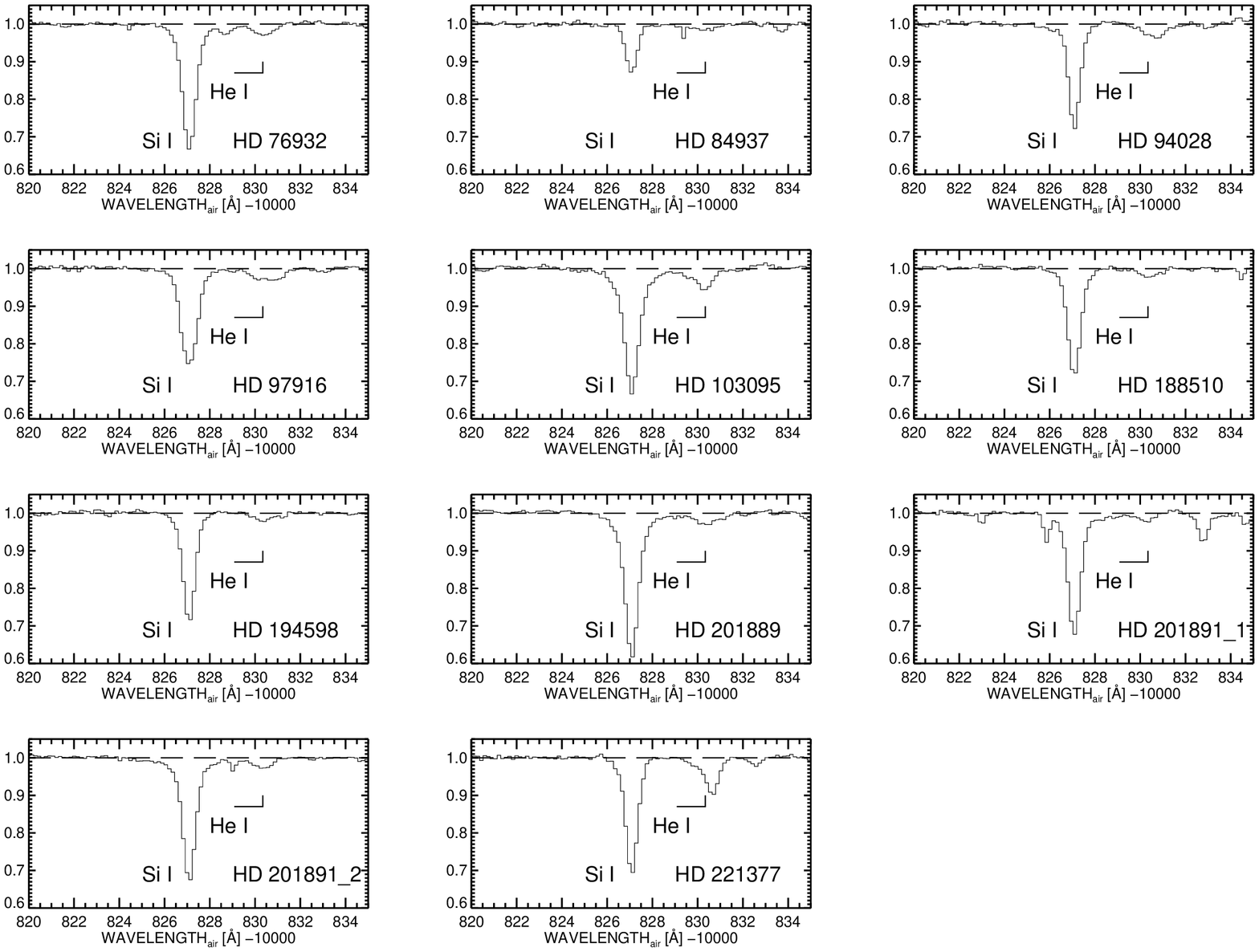}
\caption{Spectra of the $\lambda$10830 \ion{He}{1} line (normalized to a 
local continuum) for metal-poor stars listed in Table 1. Each spectrum is
plotted in the stellar rest frame.}
\end{center}
\end{figure*}

The properties of the \ion{He}{1} line that are of interest here are the
strength of the absorption plus the possibility of any profile asymmetry that
might be indicative of mass motions in the outer atmosphere. The spectra of 
the stars in our program are further presented in Figure 3(a-b) in a 
higher resolution version that shows a limited 
wavelength region (10820 to 10835 \AA) near the \ion{Si}{1} and \ion{He}{1} 
transitions. The spectra are aligned on the strong \ion{Si}{1} 10827.1 \AA\ 
transition. The location of any potential \ion{He}{1} line is marked in each 
panel together with a strong line of \ion{Si}{1}. Telluric features can be
identified by comparison with Figure 2.

The equivalent width ($EW$) of each \ion{He}{1} line was measured using the 
{\tt splot} option within the IRAF software package\footnote{IRAF is
distributed by the National Optical Astronomy Observatories, which is operated
by the Association of Universities for Research in Astronomy, Inc. under 
cooperative agreement with the National Science Foundation.} by direct 
integration of the continuum normalized profile. In instances where telluric 
water vapor intervened close to, but separable from, the \ion{He}{1} line,
deblending of the wavelength region containing 
the \ion{Si}{1}, H$_2$O, and \ion{He}{1} features was carried out by 
simultaneously fitting the profiles of all three features. 
Another complication can be the presence of a \ion{Ti}{1} line 10828.04 \AA\
in the coolest stars; by contrast the strongest helium line component is at
10830.34 \AA\ and the weakest of the triplet is at 10829.08 \AA, such that
the \ion{Ti}{1} line may blend with the \ion{He}{1} profile.

The field stars in Table 1 have a range of radial velocity. There is one 
substantial water vapor line at a terrestrial rest wavelength near 10832 \AA\ 
in the vicinity of the \ion{Si}{1} and \ion{He}{1} lines. This telluric feature
can be identified, for example, in Figure 2 between the \ion{Si}{1} and 
\ion{He}{1} lines in the spectra of HD 7424, HD 31128, HD 64606, and HD 76932. 
These stars have radial velocities of 85-121 km s$^{-1}$. 
Over the course of our 1.5-night observing run the 
\ion{He}{1} $EW$s of stars with radial velocities in the range
$v_r \sim 30$-80 km s$^{-1}$ were susceptible to overlap with the 
$\lambda$10832 telluric line. As noted above, several hot stars were observed
to measure a telluric spectrum. However, rather than
use hot-star spectra as telluric templates to divide into the object spectrum
we adopt a technique analogous to that employed by Dupree et al. (2011) to 
correct the \ion{He}{1} $EW$s for telluric absorption when necessary. There is 
another strong telluric feature in the vicinity of $\lambda$10860, and for many
of the metal-poor star spectra the equivalent width of both this feature and 
the $\lambda$10832 telluric line can be cleanly measured. Similar measurements 
were made from the hot star spectra obtained at various airmasses during
the run. Figure 4 shows that there is a correlation between the equivalent
widths of these two telluric features, and the solid line shows a linear least 
squares fit which has the equation $EW(10832) = 0.435 + 0.412 EW(10860)$. In 
the case of stars for which Figure 2 indicates that a telluric correction was 
needed to the \ion{He}{1} $EW$ measurements this correction was derived from
the $\lambda$10860 line $EW$ transformed via the relation shown in Figure 4.
The value of $EW$(10832) obtained in this way was subtracted from the measured
\ion{He}{1} $EW$. Corrections range from 41 m\AA\ for HD 23439B 
to 10 m\AA\ for HD 188510. Stars for which such a telluric correction has 
been applied are designated in Table 1. This technique has an advantage that
the telluric correction is not estimated from hot star spectra obtained at
a different time and place in the sky from the subdwarf observations.

\begin{figure}
\figurenum{4}
\begin{center}
\includegraphics[scale=0.5]{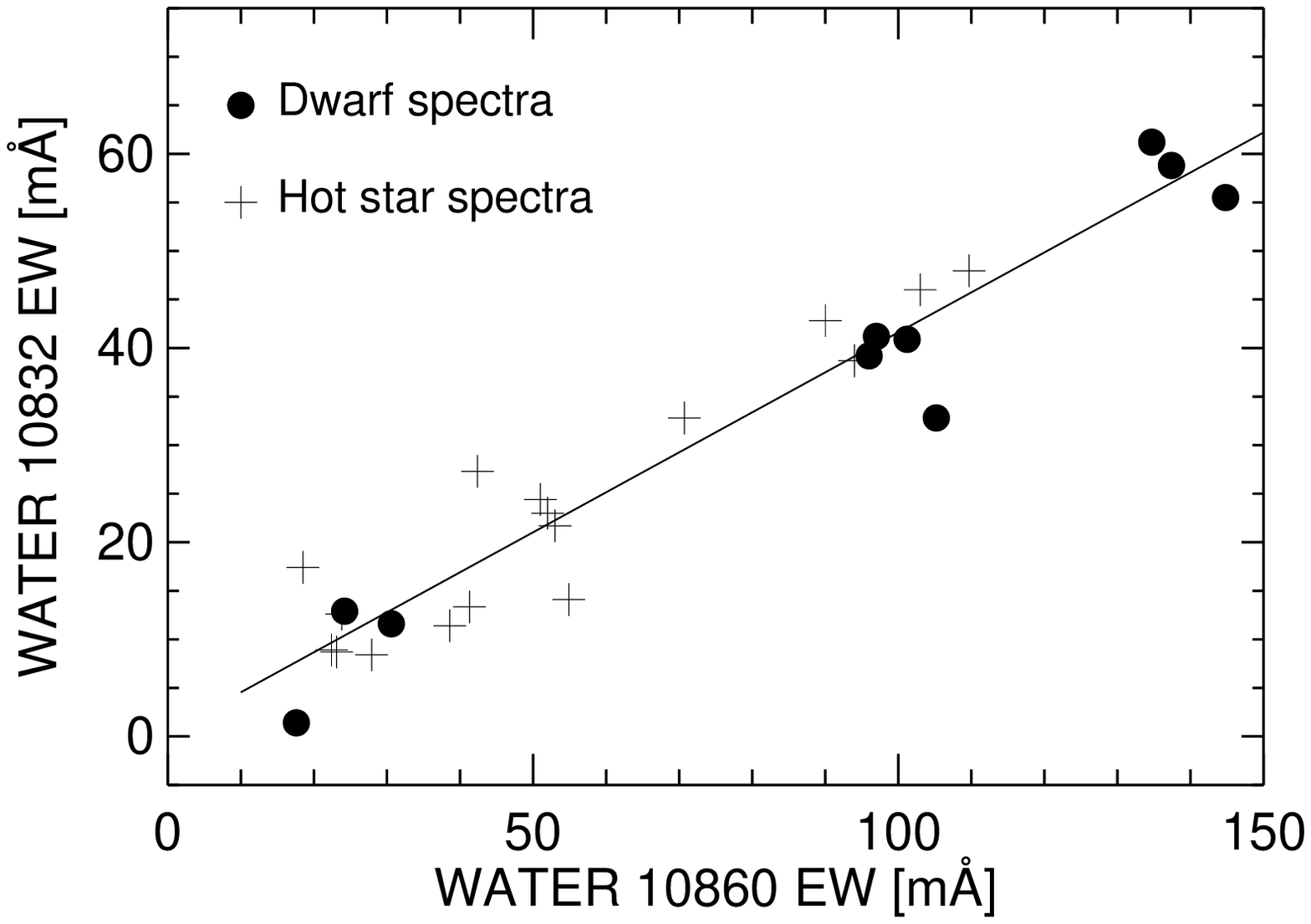}
\caption{The equivalent width of the $\lambda$10832 telluric line versus
the $EW$ for the $\lambda$10860 telluric line. Filled circles and crosses
denote measurements derived from spectra of subdwarf and hot stars 
respectively}
\end{center}
\end{figure}

For the star HD 103095, it is not the $\lambda$10832 telluric line that is
a problem but another feature near 10825 \AA. It is a very weak line and 
could only be measured cleanly in a few stars. This line was used in 
a similar manner to derive a telluric correction to the \ion{He}{1} $EW$
for HD 103095. The correction is small and amounts to 6 m\AA.

The resulting measurements of \ion{He}{1} $EW$ for each star in our program 
are listed in Table 1 and plotted against $(B-V)$ color and [Fe/H] abundance 
in Figures 5 and 6 respectively. There is little hint of any correlation 
between equivalent width and either $(B-V)$ color or [Fe/H] metallicity among 
the dwarfs. This conclusion is in accord with the findings of 
Takeda \& Takada-Hidai (2011). Equivalent widths of the \ion{He}{1} lines for 
HD 194598 and HD 201891 
were measured by Takeda \& Takada-Hidai (2011) based on observations obtained
on 2009 July 29 and 30 (UT) with the Infrared Camera and Spectrograph on the 
Subaru telescope. They found a result of $EW=30.5$ m\AA\ for HD 194598 compared
with 27 m\AA\ from our NIRSPEC observations, and 26.2 m\AA\ for HD 201891 
compared with 17 m\AA\ and 27 m\AA\ from the two NIRSPEC spectra. There is 
good consistency between the Keck and Subaru \ion{He}{1} spectra for both
of these stars. The equivalent widths for the dwarfs in the Takeda \& 
Takada-Hidai (2011) sample range from 23-80 m\AA\ with the majority of their 
dwarfs being $\sim$ 23-45 m\AA. Thus the range in \ion{He}{1} $EW$ among the 
subdwarfs in the two investigations is quite similar.

By and large the equivalent widths of the \ion{He}{1} line among the dwarfs
are comparable to those exhibited by many (but not all) of the red giants
and red horizontal branch (RHB) stars in the sample of Dupree et al. (2009). 
The largest equivalent widths ($EW > 100$ m\AA) found for some metal-poor 
giants and RHB stars by Dupree et al. (2009) are not present among the 
subdwarfs in our sample. However, many metal-poor giants in Dupree et al. 
(2009) do exhibit $\lambda$10830 equivalent widths of less than 70 m\AA, and 
overlap the range of $EW$ seen among the subdwarfs in Figure 5.

\begin{figure}
\figurenum{5}
\begin{center}
\includegraphics[angle=90,scale=0.4]{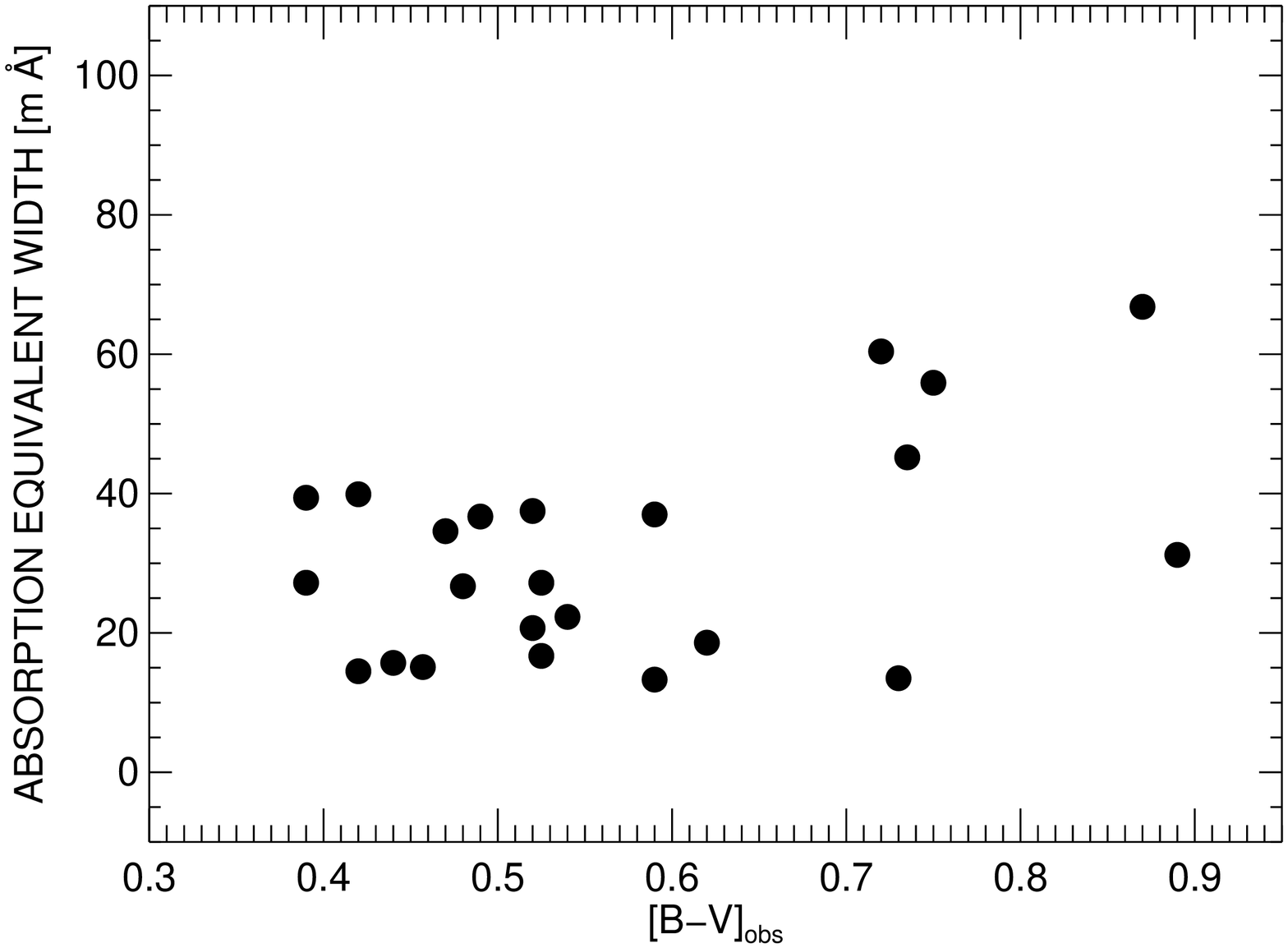}
\caption{The equivalent width of the $\lambda$10830 \ion{He}{1} line
versus $(B-V)$ color for stars listed in Table 1.}
\end{center}
\end{figure}

A comparison can also be made between the metal-poor subdwarfs and Population I
dwarfs. This is done in Figure 7 in which the \ion{He}{1} $EW$ is plotted
versus $B-V$ for the metal-poor dwarfs from Table 1 and Population I dwarfs
from two sources. The filled symbols denote subdwarfs from Table 1. The open 
circles denote the absorption equivalent widths of dwarf stars taken from 
Zarro \& Zirin (1986) where the error bar represents their estimated 
measurement error from the spectra (having a resolution of $R=11000$) for stars
with equivalent width less than 50 m\AA. The open square marks the value for 
the active dwarf $\epsilon$ Eri measured with a resolution of $R=170,000$ from 
Sanz-Forcada \& Dupree (2008). The variability is most likely real. The 
metal-poor subdwarfs have equivalent widths comparable to the weaker helium 
lines seen among dwarf stars of solar metallicity. There is a much greater 
range in $EW$ among the Population I dwarfs, many of which have considerably 
stronger \ion{He}{1} lines ($EW > 100$ m \AA) than the subdwarfs.

In searching for evidence of an outflow one is looking for an asymmetric
profile in which absorption exhibits a tail towards the \ion{Si}{1} line
at 10827 \AA\ or a profile whose greatest depth is offset from the
expected rest frame wavelength. Concerning line profile shapes, some dwarfs 
might show evidence for an asymmetric \ion{He}{1} profile, including
HD 23439B, HD 31128, HD 59374, HD 64606, HD 94028, HD 103095, and HD 221377.
HD 64606 is problematic because of a telluric feature between the \ion{He}{1}
and \ion{Si}{1} lines. HD 103095 shows a somewhat asymmetic profile but the
line center is very close to the rest frame wavelength and the degree of 
extension toward the \ion{Si}{1} line is modest.
Possible blue-extended absorption in the case of both HD 23439B
may be an artifact of incomplete continuum normalization since the line profile
seems quite symmetric and the \ion{Si}{1} line is quite broad with extensive
wings on both blue and red sides. 

\begin{figure}
\figurenum{6}
\begin{center}
\includegraphics[angle=90,scale=0.4]{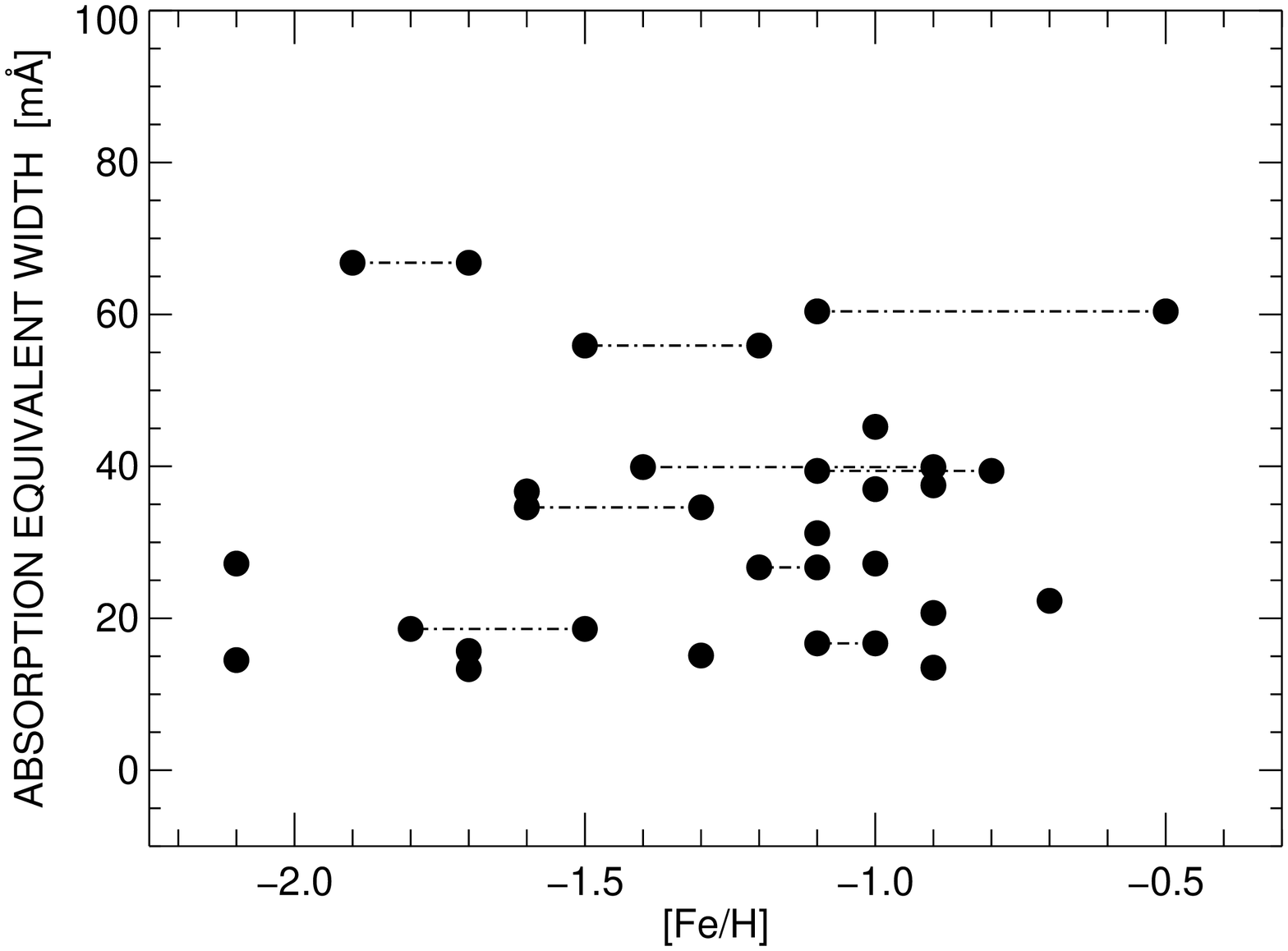}
\caption{The equivalent width of the $\lambda$10830 \ion{He}{1} line
versus [Fe/H] metallicity for stars listed in Table 1. Where two different
values of [Fe/H] are listed for a star in Table 1 a pair of points connected by
a dashed line is plotted. Slight offsets in $EW$ have been incorporated for 
clarity in display.}
\end{center}
\end{figure}

Some metal-poor dwarfs do show possible evidence of \ion{He}{1} line asymmetry 
and/or velocity shifts of the line core from the rest frame of the 
photosphere. Such cases are highlighted in column 11 of Table 1 with several 
annotations denoting either an asymmetric \ion{He}{1} profile that is 
extended on the blue or red sides, or whether the deepest part of the 
line profile is offset from the expected restframe velocity, or whether 
emission appears to be present. In none of the spectra does the blue wing of 
the \ion{He}{1} absorption line extend to the rest frame wavelength of the 
neighboring \ion{Si}{1} line. In the case of five stars noted in Table 1 
the possible presence of a telluric feature may be influencing the
appearance of the line profile.
 
Various figures in Dupree et al. (1992), Smith et al. (2004), and  Dupree 
et al. (2009) show examples of the types of fast outflow \ion{He}{1} profiles 
that have been discovered among some Population II red giants. In such stars 
the broad \ion{He}{1} absorption profile can extend blueward into the region 
of the 10827.1 \AA\ \ion{Si}{1} line. None of the subdwarf 
spectra in Figure 3 show the type of extended blue wing absorption seen 
in the field giants HD 6833 (Dupree et al. 1992) or HD 122563 (Smith et al. 
2004) for example. As such, the metal-poor dwarfs do not show evidence for the 
type of high-speed outflows found among such metal-poor giants as HD 6833 
(Dupree et al. 1992). Note this does not necessarily indicate that high speed 
flows are absent in these stars. High-velocity absorption arises in giants 
because the population in the metastable level spans an extended chromosphere 
allowing scattering of near-infrared photons and tracing the acceleration in 
the outflow. In contrast, the scale height of the atmosphere is less in dwarf 
stars than giants and helium would be present over a more narrow atmospheric 
region where the outflow velocity may be lower or nearly constant.

\section{Comments on Individual Stars}

Our general results have been presented in the preceding section. In this 
section comments about individual stars are given.

HD 7424. There is a difference of 0.6 dex in the [Fe/H] measurements of
Peterson (1981) and Tomkin et al. (1992). The \ion{He}{1} line may be
blended with another very weak feature.

HD 59984 is considered to be a metal-poor disk dwarf and has been the subject
of many metallicity measurements in addition to the work of Chen et al. (2000).
The majority of the published values of [Fe/H] are close to $-0.7$ to $-0.8$ 
dex. It has a \ion{He}{1} $EW$ that is consistent with that of more metal-poor 
dwarfs of similar $(B-V)$ color that belong to the halo.

HD 64606 is listed as a spectroscopic binary in the catalog of Pourbaix et al.
(2004) with a period of about 450 days. It was observed to exhibit \ion{Ca}{2}
H and K emission by Smith \& Churchill (1998) with an asymmetry
parameter of $V/R > 1$, which is typical of other subdwarfs in their survey.
HD 64606 was included in a search for X-ray activity among Population II field
binaries by Ottman et al. (1997) using the {\it ROSAT} all-sky survey. The
observations provided only an upper limit on any X-ray luminosity, this star
was not among those Population II binaries detected by {\it ROSAT}.
The \ion{He}{1} line is relatively weak.

HD 23439B, HD 59374, HD 59984, HD 94028 and HD 97916. The possible asymmetries 
and/or wavelength offsets of the \ion{He}{1} line noted in Table 1 for these 
stars may be influenced by the presence of a telluric feature.

\begin{figure}
\figurenum{7}
\begin{center}
\includegraphics[angle=90,scale=0.4]{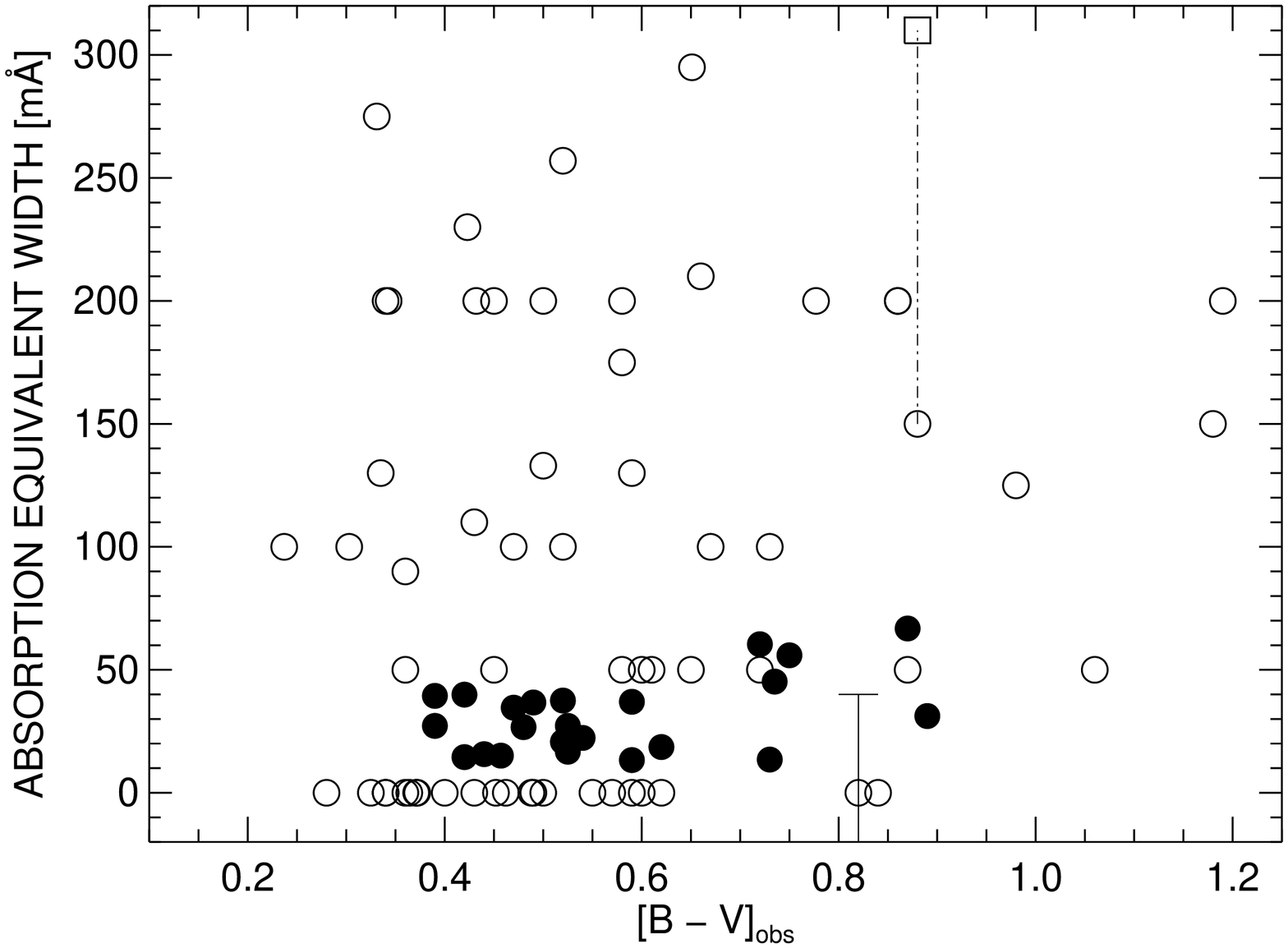}
\caption{The equivalent width of the $\lambda$10830 \ion{He}{1} line
versus $(B-V)$ color for metal-poor subdwarfs stars listed in Table 1 
(filled circles), as well as Population I dwarfs (open circles) from 
Zarro \& Zirin (1986). The open square denotes a measurement for the star
$\epsilon$ Eri obtained by Sanz-Forcada \& Dupree et al. (2008).}
\end{center}
\end{figure}

\section{Discussion}

The results of this work can be summarized as follows:\\
(i) Metal-poor field subdwarfs and dwarfs commonly exhibit a \ion{He}{1}
absorption line indicative of the presence of chromospheres.\\
(ii) The equivalent width of the \ion{He}{1} absorption line is typically less
than 70 m\AA\ among metal-poor dwarfs. Our results are consistent 
with the observations of Takeda \& Takada-Hidai (2011) in this regard.
The range of \ion{He}{1} equivalent widths among the subdwarfs overlaps the 
range displayed by metal-poor red giants, however, some metal-poor giants do 
show notably stronger $\lambda$10830 lines.\\
(iii) There is no evidence in our data set for any correlation between
\ion{He}{1} line equivalent width and either [Fe/H] metallicity or 
$(B-V)$ color among metal-poor dwarf stars.\\
(iv) The dwarfs stars in our sample lack the strong ${\rm EW} > 100$ m\AA\ 
absorption lines evident among some field RHB stars.\\ 
(v) Whereas some asymmetric profiles are discernible among metal-poor dwarfs, 
they generally do not display the type of high-speed blue-shifted outflow 
absorption found by Dupree et al. (1992, 2009) among some metal-poor red 
giants.

The process(es) responsible for populating the lower state of the 
10830 \AA\ triplet line have been the subject of debate (e.g., Zirin 1976, 
Cuntz \& Luttermoser 1990; Sanz-Forcada \& Dupree 2008).
Among Population I dwarfs later than spectral type $\sim$ F7 there is a 
correlation between soft X-ray flux and the equivalent width of the \ion{He}{1}
$\lambda$10830 line (Zarro \& Zirin 1986; Takeda \& Takada-Hidai 2011).
This correlation has been documented down to X-ray flux levels of
$\log (f_{\rm x}/f_{\rm bol}) \sim -6.2$ at which the $\lambda$10830 equivalent
width falls within the range $1.4 < \log EW < 2.0$ (see Fig.~5 of Takeda \& 
Takada-Hidai 2011). Furthermore in the case of the Sun there are spatial 
correlations between the $\lambda$10830 line strength and X-ray and EUV flux 
(Harvey \& Sheeley 1977; Thompson et al. 1993; Braj\v{s}a et al. 1996). 
These correlations have been interpreted as resulting from photoionization
of \ion{He}{1} by X-rays followed by recombination to the triplet levels of 
neutral helium (e.g., Zirin 1975).

The most metal-poor halo dwarfs in Table 1 are presumably among the oldest 
stars in the Galaxy. As such their coronal activity is expected to have 
decreased to low levels. Such stars are expected to be 
considerably older than the Population I dwarfs in the Zarro \& Zirin (1986) 
survey. Yet one can point to examples of metal-poor subdwarfs in Table 1 that 
have comparable \ion{He}{1} $EW$s to some dwarfs in the Zarro \& Zirin (1986) 
sample (Figure 7). Does this imply a decoupling of the \ion{He}{1}
line strength and coronal flux among dwarf stars of low levels of activity?
Takeda \& Takada-Hidai (2011) argued that the presence of \ion{He}{1} 
equivalent widths of $1.0 < \log EW < 2.0$ among subdwarfs is consistent with
old dwarf stars attaining a base level of activity that is driven by some
heating mechanism that is unrelated to a magnetic dynamo. Since the \ion{He}{1}
$\lambda$10830 lower level is still being excited among Population II 
subdwarfs, if radiative excitation from high-energy coronal photons is the
relevant mechanism then it might be expected that such stars should be 
detected in soft X-rays if deep enough integrations can be acquired by using 
orbiting X-ray observatories. However, none of the stars in Table 1 are found 
as detections in the {\it ROSAT All-Sky Survey Faint Source Catalogue} (Voges 
et al. 2000).\footnote{Stellar metallicity might play a role here. Perhaps
for a given energy input metal-poor atmospheres might be warmer than 
chromospheres of solar metallicity since they may not radiate as efficiently. 
If so, then higher chromospheric temperatures would increase the strength of 
the helium line.} 

The NIRSPEC spectra do not appear to have revealed any evidence among the
subdwarfs in Table 1 for the fast $\sim 100$ km s$^{-1}$ outflows seen by
Dupree et al. (1992, 2000) among some Population II red giants. This at least
adds support to the suggestion that the fast winds seen in the \ion{He}{1}
profile are a by-product of red giant evolutionary processes. It is appropriate
here to distinguish between an outflow (that is, a motion of material away from
a star) and a wind (material exceeding the stellar escape velocity at a given 
radius). A notable distinction between the metal-poor red giants and the 
subdwarfs is not just the velocity of the outflow, but also that the subdwarfs 
have larger escape velocities. The data in this paper give no evidence
that such outflowing material as is observed is unbound from the star, unlike 
the case for some red giants discussed by Dupree et al. (2009).

Among Population I giants of late-K spectral type O'Brien \& Lambert 
(1986) and Lambert (1987) found evidence from the $\lambda$10830 profiles for 
time-variable mass motions in the chromosphere suggestive of ``episodic 
ejection of matter and a subsequent return of some fraction of this matter to 
the photospheres in these stars.'' By contrast early-K Population I giants 
were found by them to have constant $\lambda$10830 profiles not dissimilar to 
that of the Sun. In the case of $\alpha$ Boo Lambert (1987) concluded that the
profile variations may be periodic. Evidently when both Population I and II
red giants evolve into the K spectral class dynamic phenomena become
evident in the outer atmosphere. The challenge is to determine whether these 
outflows seen in the \ion{He}{1} line are regular, semi-regular, or irregular,
what the driving mechanism of the outflows is, and whether these mass motions
trace significant episodes of mass loss, as argued by Dupree et al. (2009).
 
\acknowledgements

The authors wish to recognize and acknowledge the very significant cultural 
role and reverence that the summit of Mauna Kea has always had within the 
indigenous Hawaiian community.  We are most fortunate to have the opportunity 
to conduct observations from this mountain. We thank the referee for useful 
comments on the paper.

\clearpage

\begin{deluxetable}{llrcrrccrrl}
\tabletypesize{\small}
\tablecolumns{11}
\tablewidth{0pt}
\tablenum{1}
\tablecaption{\ion{He}{1} Line Characteristics of Metal-Poor Stars} 
\tablehead{
\colhead{HD}      &        
\colhead{Exp}     &                            
\colhead{$V$}     &  
\colhead{$B-V$}   &      
\colhead{Par}     & 
\colhead{$M_V$}   &    
\colhead{[Fe/H]}  &       
\colhead{[Fe/H]}  &
\colhead{$v_r$}   &
\colhead{$EW$(He)}  & 
\colhead{Notes}   \\
\colhead{(1)}     &       
\colhead{sec}     &  
\colhead{(3)}     & 
\colhead{(4)}     &   
\colhead{mas}     &  
\colhead{(6)}     &  
\colhead{(7)}     & 
\colhead{Ref}     &
\colhead{km s$^{-1}$} &   
\colhead{m\AA}   & 
\colhead{(11)}
}
\startdata
   3179\tablenotemark{a}     & 2$\times$1$\times$180      &  9.75  & 0.74 &   3.47  &   2.45  & --1.0      & a  & --74  &  45.2  &   \\  
   3567     & 2$\times$1$\times$90       &  9.25  & 0.46 &   9.51  &   4.15  & --1.3        & c    & --50  &  15.1  &       \\
   7424     & 2$\times$1$\times$180      & 10.08  & 0.72 &   6.15  &   4.02  & --0.5,--1.1  & d,e  &  +85  &  60.4  &       \\   
   16031\tablenotemark{b}    & 2$\times$1$\times$120      &  9.78  & 0.44 &   8.93  &   4.53  & --1.7        & c    &  +24  &  15.7  &  o?   \\
   23439B\tablenotemark{b}   & 2$\times$1$\times$90       &  8.77  & 0.89 &  40.8   &   6.82  & --1.1        & f    &  +50  &  31.2  &  o?   \\
   25329    & 2$\times$1$\times$45       &  8.50  & 0.87 &  54.68  &   7.19  & --1.7,--1.9  & b,c  & --30  &  66.8  &       \\     
   31128    & 2$\times$2$\times$(60+120) &  9.13  & 0.49 &  15.0   &   5.01  & --1.6        & c    & +105  &  36.7  &  o?   \\
   59374\tablenotemark{b}    & 2$\times$2$\times$60       &  8.49  & 0.52 &  20.2   &   5.02  & --0.9        & c    &  +80  &  37.5  &  ar,o \\ 
   59984\tablenotemark{b}    & 2$\times$1$\times$60       &  5.90  & 0.54 &  35.82  &   3.67  & --0.7        & h    &  +55  &  22.3  &  ab,o  \\
   64090    & 2$\times$2$\times$60       &  8.30  & 0.62 &  34.3   &   5.98  & --1.5,--1.8  & b,c  &  --240 &  18.6  &       \\    
   64606\tablenotemark{b}    & 2$\times$2$\times$15       &  7.44  & 0.73 &  49.78  &   5.93  & --0.9        & c    &  +93   &  13.5  &       \\    
   74000    & 2$\times$2$\times$180      &  9.67  & 0.42 &   7.86  &   4.15  & --2.1        & c    &  +204  &  14.5  &       \\  
   76932\tablenotemark{b}    & 2$\times$2$\times$(5+5)    &  5.82  & 0.52 &  47.54  &   4.21  & --0.9        & c    &  +121  &  20.7  &       \\  
   84937    & 2$\times$2$\times$45       &  8.32  & 0.39 &  13.74  &   4.01  & --2.1        & c    &  --17  &  27.2  &       \\  
   94028\tablenotemark{b}    & 2$\times$1$\times$100      &  8.23  & 0.47 &  21.11  &   4.85  & --1.3,--1.6  & i,c  &  +62  &  34.6  &  ab,o? \\
   97916\tablenotemark{b}    & 2$\times$4$\times$60       &  9.21  & 0.42 &   8.95  &   3.97  & --0.9,--1.4  & c,j  &  +55  &  39.9  &  ar   \\
   103095\tablenotemark{b}   & 2$\times$2$\times$15       &  6.45  & 0.75 & 109.99  &   6.66  & --1.2,--1.5  & b,c  &  --98  &  55.9  &  ab   \\
   188510\tablenotemark{b}   & 2$\times$4$\times$60       &  8.83  & 0.59 &  26.71  &   5.96  & --1.7        & c    &  --193  &  13.3  &       \\  
   194598   & 4$\times$2$\times$120      &  8.34  & 0.48 &  17.00  &   4.49  & --1.1,--1.2  & b,c  &  --246  &  26.7  &       \\
   201889   & 2$\times$3$\times$45       &  8.06  & 0.59 &  18.60  &   4.41  & --1.0        & c    &  --103  &  37.0  &       \\
   201891-1 & 2$\times$1$\times$45       &  7.37  & 0.53 &  29.10  &   4.69  & --1.0,--1.1  & b,c  &   --45  &  16.7  &  ab?  \\ 
   201891-2 & 2$\times$3$\times$25       &  ....  & .... &  .....  &   ....  & ...          &      &  ....   & 27.2  &  ab?  \\
   221377\tablenotemark{b}   & 2$\times$1$\times$45       &  7.57  & 0.39 &  13.10  &   3.16  & --0.8,--1.1  & c,k  &  +27  &  39.4  &  ab,o \\
\enddata
\tablenotetext{a}{Possibly a subgiant star on basis of position in the 
color-magnitude diagram of Figure 1}
\tablenotetext{b}{The value of $EW$(He) have been corrected for a telluric feature 
as discussed in the text.}

\end{deluxetable}

\noindent Reference key (column 8) for tabulated values of [Fe/H]:\\
(a) Burris et al. (2000)\\
(b) Gratton et al. (2000)\\
(c) Fulbright (2000)\\
(d) Peterson (1981)\\
(e) Tomkin et al. (1992)\\
(f) Tomkin \& Lambert (1999)\\
(g) For \& Sneden (2010)\\
(h) Chen et al. (2000)\\
(i) Clementini et al. (1999)\\
(j) Pilachowski et al. (1993)\\
(k) Abia et al. (1988)

\noindent Annotations in column 11: ab = a star with an asymmetric \ion{He}{1} 
profile that is extended on the blue side; ar = a star with an asymmetric 
\ion{He}{1} profile that is extended on the red side; o = a case where the 
deepest part of the line profile is offset from the expected restframe 
velocity; and em = emission.

\clearpage

\end{document}